\documentclass[11pt, a4paper]{article}
\usepackage{jheparxiv}
\usepackage[utf8]{inputenc}
\usepackage{amsmath}
\usepackage{amsfonts}
\usepackage{amssymb}
\usepackage{latexsym}
\usepackage{mathrsfs}
\usepackage{braket}		
\usepackage{graphicx}
\usepackage{color}
\usepackage{xcolor}
\usepackage{slashed}
\usepackage{twistor}
\usepackage[all]{xy}

\newcommand{\sa}{\mathsf{a}}

\newcommand{\sA}{\mathsf{A}}

\newcommand{\sG}{\mathsf{G}}
\newcommand{\sH}{\mathsf{H}}

\renewcommand{\d}{\mathrm{d}}
\renewcommand{\sb}{\mathsf{b}}
\renewcommand{\sc}{\mathsf{c}}

\subheader{\hfill \texttt{IMPERIAL-TP-TA-2018-02}}

\title{Yang-Mills theory from the worldsheet}

\author[a]{Tim Adamo,}
\author[b]{Eduardo Casali}
\author[b]{\& Stefan Nekovar}

\affiliation[a]{Theoretical Physics Group, Blackett Laboratory \\
        Imperial College London, SW7 2AZ, United Kingdom}

\affiliation[b]{The Mathematical Institute \\
        University of Oxford, Woodstock Road, OX2 6GG, United Kingdom}

\emailAdd{t.adamo@imperial.ac.uk}
\emailAdd{[casali,nekovar]@maths.ox.ac.uk}

\abstract{We give a new description of classical Yang-Mills theory by coupling a two-dimensional chiral CFT (which gives the tree-level S-matrix of Yang-Mills theory at genus zero) to a background non-abelian gauge field. The resulting model is solvable, and when the gravitational degrees of freedom are decoupled the non-linear Yang-Mills equations are imposed as an exact worldsheet anomaly cancellation condition. When gravitational modes are reinstated, we find gauge anomalies analogous to those appearing in heterotic string theory.}

\begin{document}
\notoc

\maketitle\vfill

\section{Introduction}

The equations of motion for a classical field theory are usually understood as the Euler-Lagrange equations of a corresponding action functional. For certain theories, such as gauge theory and gravity, the equations of motion can also famously be derived from a low-energy expansion of string theory~\cite{Callan:1985ia,Abouelsaood:1986gd,Banks:1986fu}. Coupling the Polyakov action to background gauge or gravitational fields leads to a conformal anomaly. Since the resulting worldsheet action is a complicated interacting 2d conformal field theory (CFT), this anomaly can only be computed perturbatively in a small parameter, taken to be the inverse string tension. To lowest order in this parameter, anomaly cancellation imposes the field equations of gauge theory or gravity on the background fields; the higher-order corrections impose an infinite tower of additional higher-derivative equations. 

Recent progress in the study of scattering amplitudes has suggested a new description of field theories as certain string theories in their own right. Compact formulae for the full tree-level S-matrix of many massless QFTs (including gauge theory and gravity) present scattering amplitudes in terms of localized integrals on the Riemann sphere rather than a sum of Feynman diagrams or the low-energy limit of an ordinary string theory amplitude~\cite{Cachazo:2013hca,Cachazo:2014xea}. These formulae arise as the genus zero correlation functions of certain chiral 2d CFTs known as \emph{ambitwistor strings}~\cite{Mason:2013sva}, which have a finite massless spectrum and no tunable parameter on the worldsheet. A natural question is therefore: can ambitwistor strings be coupled to background fields to give a non-linear description of the underlying field theories? Unlike ordinary string theory, such a description should be exact -- that is, computable without recourse to an infinite perturbative expansion.

For the case of the NS-NS supergravity (consisting of a graviton, dilaton and two-form $B$-field) this was answered in the affirmative: the NS-NS sector of supergravity is described \emph{exactly} by a worldsheet theory at the level of its non-linear equations of motion~\cite{Adamo:2014wea}. More recently, it was shown that the abelian Maxwell equations could also be obtained in a similar fashion~\cite{Adamo:2017sze}. However, this success has not been extended to other field theories due to a variety of subtleties associated with coupling to background fields and non-unitary gravitational modes which do not exist in the case of NS-NS supergravity.

In this paper, we extend the exact worldsheet description of classical field theories to include Yang-Mills theory. This is accomplished by coupling a \emph{heterotic} version of the ambitwistor string to a non-abelian background gauge field.\footnote{In~\cite{Azevedo:2016zod} a form of the heterotic ambitwistor string with background fields was also studied, but only classically on the worldsheet.} The model contains both gauge theoretic and (non-unitary) gravitational degrees of freedom, but the latter can be locally decoupled on the worldsheet. Gauge fixing the worldsheet action leads to potential anomalies; remarkably, the \emph{only} conditions imposed on the background gauge field by anomaly cancellation are the (non-linear) Yang-Mills equations. We also show that re-coupling the gravitational modes leads to gauge anomalies, analogous to but distinct from the well-known anomaly~\cite{Green:1984sg} of the standard heterotic string.


\section{The worldsheet model}

The heterotic ambitwistor string has worldsheet action~\cite{Mason:2013sva}:
\be\label{wsa1}
S=\frac{1}{2\,\pi}\int_{\Sigma} P_{\mu}\,\dbar X^{\mu}+\frac{1}{2}\,\psi_{\mu}\,\dbar\psi^{\mu} +S_{C}\,,
\ee
where $\Sigma$ is a closed Riemann surface, $P_{\mu}$ has conformal weight $(1,0)$ on $\Sigma$, $\psi^{\mu}$ is a real fermion with conformal weight $(\frac{1}{2},0)$, and $S_C$ is the action of a holomorphic worldsheet current algebra for some gauge group $G$ (assumed to be simple and compact). The current, $j^{a}$, associated to $S_C$ has conformal weight $(1,0)$ and its OPE on $\Sigma$ takes the form:
\be\label{calg}
j^{\sa}(z)\,j^{\mathsf{b}}(w)\sim \frac{k\,\delta^{\mathsf{ab}}}{(z-w)^2} + \frac{f^{\mathsf{abc}}\,j^{\mathsf{c}}(w)}{z-w}\,,
\ee
where the indices $\mathsf{a}, \mathsf{b},\ldots$ run over the adjoint representation of $G$, $k$ is the level of the worldsheet current algebra, and $f^{\mathsf{abc}}$ are the structure constants.

It is easy to see that \eqref{wsa1} is invariant under global holomorphic reparametrizations as well as gauge transformations generated by the constraints $P^2=0=\psi\cdot P$. These symmetries can be gauged, resulting in a consistent, anomaly-free 2d CFT when the space-time dimension $d$ and central charge of the worldsheet current algebra $\mathfrak{c}$ are related by: $\mathfrak{c}=41-\frac{5d}{2}$.

The spectrum of this model contains gluons of the gauge group $G$ as well as gravitational degrees of freedom. The gravitational modes correspond to a non-unitary theory of gravity which is fourth-order in derivatives~\cite{Azevedo:2017lkz}. The genus zero worldsheet correlation functions of gluon vertex operators generate the entire tree-level S-matrix of Yang-Mills theory in the scattering equations form of~\cite{Cachazo:2013hca} by isolating the single trace contributions to the correlator from the worldsheet current algebra. Double (and higher) trace terms -- which contribute with higher powers of the level $k$ -- are mediated by the non-unitary gravitational modes.

As first observed in the context of similar chiral heterotic-like worldsheet models~\cite{Berkovits:2004jj}, these `bad' gravitational modes can be decoupled at genus zero by taking a limit where $k\rightarrow 0$ while $\frac{k}{g^2_s}$ is held fixed, for $g_s$ the `string' coupling constant which effectively counts the genus of $\Sigma$. In this limit, only single trace contributions to a worldsheet correlation function survive at genus zero. Globality and unitarity of the worldsheet current algebra dictate that $k$ be a positive integer, so the $k\rightarrow 0$ limit must be viewed as a purely formal one which effectively removes the second order pole from the OPE \eqref{calg}. Since our primary concern will be anomalies, which can be computed \emph{locally} on $\Sigma$, the formality of the limit will not be a problem. 

Fixing $k\rightarrow 0$, introduce a background gauge potential $A^{\sa}_{\mu}(X)$ valued in the adjoint of $G$, which couples to the worldsheet current algebra by
\be\label{back1}
S_{C}\rightarrow S_{C}+\frac{1}{2\,\pi}\int_{\Sigma}A^{\sa}_{\mu}\,j^{\sa}\,\dbar X^{\mu}\,.
\ee
The purely quadratic nature of the worldsheet model \eqref{wsa1} can be preserved by absorbing the explicit dependence on the background into the conjugate of $X^{\mu}$ (cf., \cite{Adamo:2014wea} for the analogous procedure in the gravitational context):
\be\label{pidef}
P_{\mu}\rightarrow P_{\mu} +A^{\sa}_{\mu}\,j^{\sa}:=\Pi_{\mu}\,.
\ee
This leads to a worldsheet action
\be\label{wsa2}
S=\frac{1}{2\,\pi}\int_{\Sigma}\Pi_{\mu}\,\dbar X^{\mu}+\frac{1}{2}\,\psi_{\mu}\,\dbar\psi^{\mu} +S_{C}\,,
\ee
with free OPEs of the worldsheet fields (in the $k\rightarrow 0$ limit):
\be\label{OPEs}
X^{\mu}(z)\,\Pi_{\nu}(w)\sim \frac{\delta^{\mu}_{\nu}}{z-w}\,, \qquad \psi^{\mu}(z)\,\psi^{\nu}(w)\sim \frac{\eta^{\mu\nu}}{z-w}\,, \qquad j^{\sa}(z)\,j^{\mathsf{b}}(w)\sim \frac{f^{\mathsf{abc}}\,j^{\mathsf{c}}(w)}{z-w}\,,
\ee
for $\eta_{\mu\nu}$ is the $d$-dimensional Minkowski metric.

The price for this simplicity (which is in stark contrast to the complicated interacting 2d CFT obtained by coupling the ordinary heterotic string to a background) is that $\Pi_{\mu}$ is not invariant under local gauge transformations. Under an infinitesimal gauge transformation with parameter $\varepsilon^{\sa}$,
\be\label{gt1}
A^{\sa}_{\mu}\rightarrow f^{\mathsf{abc}}\,\varepsilon^{\mathsf{b}}\,A^{\mathsf{c}}_{\mu}-\partial_{\mu}\varepsilon^{\sa}\,, \qquad j^{\sa}\rightarrow f^{\mathsf{abc}}\,\varepsilon^{b}\,j^{\mathsf{c}}\,,
\ee
which indicates that 
\be\label{gt2}
\Pi_{\mu}\rightarrow \delta\Pi_{\mu}:=-j^{\sa}\,\partial_{\mu}\varepsilon^{\sa}\,.
\ee
This is a common feature of all curved $\beta\gamma$-systems, of which \eqref{wsa2} is an example~\cite{Nekrasov:2005wg}, and is problematic only if $\Pi_{\mu}$ has a singular OPE with itself after a gauge transformation. Fortunately, it is easy to check that
\be\label{gt3}
(\Pi_{\mu}+\delta\Pi_{\mu})(z)\,(\Pi_{\nu}+\delta\Pi_{\nu})(w)\sim 0\,,
\ee
so the structure of the OPEs \eqref{OPEs} is preserved under gauge transformations. Note that infinitesimal gauge transformations can be implemented by a local operator $\cO_{\varepsilon}=-\varepsilon^{\sa} j^{\sa}$, which obeys
\be\label{gtop}
\cO_{\varepsilon}(z)\,\cO_{\lambda}(w)\sim\frac{f^{\mathsf{abc}}\varepsilon^{\sa}\,\lambda^{\mathsf{b}}\,j^{\mathsf{c}}}{z-w}=\frac{\cO_{[\lambda,\varepsilon]}}{z-w}\,,
\ee
and acts correctly on all worldsheet fields.


\section{Yang-Mills equations as an anomaly}

The worldsheet action \eqref{wsa2} has additional symmetries beyond holomorphic reparametrization invariance. Indeed, the action is invariant under the transformations
\be\label{ftran1}
\delta X^{\mu}=-\epsilon\,\psi^{\mu}\,, \qquad \delta\psi^{\mu}=\epsilon\left(\Pi^{\mu}-A^{\mu\,\sa}j^{\sa}\right)\,, \qquad \delta\Pi_{\mu}=\epsilon \psi^{\nu}\partial_{\mu}A^{\sa}_{\nu} j^{\sa}\,,
\ee
where $\epsilon$ is a constant fermionic parameter of conformal weight $(-\frac{1}{2},0)$. These transformations are generated by a fermionic current
\be\label{Gcurr}
\sG=\psi^{\mu}\left(\Pi_{\mu}-A^{\sa}_{\mu}\,j^{\sa}\right)\,,
\ee
on the worldsheet, which is the extension of $\psi\cdot P$ to the non-trivial gauge background. This current is a holomorphic conformal primary of dimension $3/2$ and is invariant under gauge transformations of the background field.

The OPE of $\sG$ with itself has only simple poles, generating a bosonic current:
\be\label{Hcurr0}
\sG(z)\,\sG(w)\sim \frac{\sH}{z-w}\,,
\ee
where $\sH$ is the extension of $P^2$ to the non-trivial gauge background:
\be\label{Hcurr}
\sH = \Pi^2 - 2\, \Pi^\mu A_{\mu}^{\sa} j^\sa + \sA_\mu^\sa \sA^{\mu \sb} j^\sa j^\sb + \psi^\mu \psi^\nu F_{\mu\nu}^\sa j^\sa - \partial\left( \partial_\mu \sA^{\mu \sa} j^\sa \right) + f^{\sa\sb\sc} j^\sc \sA^{\mu \sb} \partial \sA_\mu^\sa\,.
\ee
As both $\sG$ and $\sH$ are composite operators on the worldsheet, their definition as currents in the fully quantum mechanical regime requires normal ordering to remove singular self-contractions. We use a point-splitting prescription to do this; for example, the explicit normal ordering of the second term in $\sH$ is given by:
\be\label{no1}
- 2\, \Pi^\mu A_{\mu}^{\sa} j^\sa(z):=\frac{\im}{\pi}\,\oint \d w\, \frac{\Pi^{\mu}(w)\,A_{\mu}^{\sa}(z) j^{\sa}(z)}{z-w}\,,
\ee
where the integral is taken on a small contour in $w$ around $z$. From now on, we assume this normal ordering implicitly. It is straightforward to check that the normal-ordered current $\sH$ is a conformal primary of dimension $2$ and gauge invariant.

Gauging the symmetries associated with these currents leads to a worldsheet action
\be\label{wsa3}
S=\frac{1}{2\,\pi}\int_{\Sigma}\Pi_{\mu}\,\dbar X^{\mu}+\frac{1}{2}\,\psi_{\mu}\,\dbar\psi^{\mu} +\chi\,\sG +e\,\sH +S_{C}\,,
\ee
where the gauge fields $\chi$, $e$ are fermionic of conformal weight $(-\frac{1}{2},1)$ and bosonic of conformal weight $(-1,1)$, respectively. The worldsheet symmetries associated with $\sG,\sH$ can now be gauge-fixed, along with holomorphic reparametrization invariance; choosing conformal gauge with $\chi=e=0$ leads to a gauge-fixed action
\be\label{wsa4}
S=\frac{1}{2\,\pi}\int_{\Sigma}\Pi_{\mu}\,\dbar X^{\mu}+\frac{1}{2}\,\psi_{\mu}\,\dbar\psi^{\mu} +b\,\dbar c + \tilde{b}\,\dbar \tilde{c} +\beta\,\dbar\gamma +S_{C}\,,
\ee
and associated BRST charge
\be\label{BRST}
Q=\oint c\,T+\gamma\,\sG + \tilde{c}\,\sH +\frac{\tilde{b}}{2}\,\gamma^2\,.
\ee
Here, $(c,b)$ and $(\tilde{c},\tilde{b})$ are fermionic ghost systems for which $c,\tilde{c}$ have conformal weight $(-1,0)$, while $(\gamma,\beta)$ are a bosonic ghost system for which $\gamma$ has conformal weight $(-\frac{1}{2},0)$. In the BRST charge, $T$ denotes the (appropriately normal-ordered) holomorphic stress tensor of the worldsheet CFT, including all contributions from the worldsheet current algebra.

This gauge fixing is anomaly-free if and only if the BRST charge is nilpotent: $Q^2=0$. Given the free OPEs \eqref{OPEs}, this calculation can be performed \emph{exactly} -- there is no need for a background field expansion as in the analogous calculation for the heterotic string~\cite{Callan:1985ia}. It is straightforward to show that $Q^2=0$ only if the central charge of the worldsheet current algebra obeys $\mathfrak{c}=41-\frac{5d}{2}$, and
\be\label{anom1}
\sG(z)\,\sH(w)\sim 0\,.
\ee
The first of these conditions kills the holomorphic conformal anomaly; remarkably, it is completely independent of the background fields. The second condition -- that the OPE between $\sG$ and $\sH$ be non-singular -- is trivially satisfied in a flat background, but becomes non-trivial in the presence of $A^{\sa}_{\mu}$. Making use of the identity
\be\label{nocurr}
j^{\sa}j^{\sb}(z) - j^{\sb}j^{\sa}(z)=f^{\mathsf{abc}}\,\partial j^{\mathsf{c}}(z)\,,
\ee
for normal-ordered products of the worldsheet current, one finds:
\be\label{anom2}
\sG(z)\,\sH(w)\sim -3 \frac{\psi^{\nu} j^{\sa}\,D^{\mu}F^{\sa}_{\mu\nu}}{(z-w)^2}-\frac{\partial\left(\psi^{\nu}j^{\sa}\,D^{\mu}F^{\sa}_{\mu\nu}\right)}{z-w}-\frac{\psi^{\mu}\psi^{\nu}\psi^{\sigma}j^{\sa}\,D_{\mu}F^{\sa}_{\nu\sigma}}{z-w}\,,
\ee
where $D_{\mu}$ is the gauge-covariant derivative with respect to $A_{\mu}^{\sa}$.

Requiring \eqref{anom1} to hold then imposes the constraints
\be\label{eoms}
D^{\mu}F^{\sa}_{\mu\nu}=0\,, \qquad D_{[\mu}F^{\sa}_{\nu\sigma]}=0\,,
\ee
on the background gauge field, which are precisely the Yang-Mills equations and Bianchi identity. Furthermore, these are the \emph{only} constraints placed on the background gauge field by anomaly cancellation in the 2d worldsheet theory. Thus, non-linear Yang-Mills theory (in any dimension) is described by an exact anomaly in a 2d chiral CFT.


\section{Recoupling gravity and the gauge anomaly}

Gravitational degrees of freedom (in the guise of multi-trace terms) can be recoupled by reinstating the level $k$ (now assumed to be a positive integer). In the ordinary heterotic string, the interplay between gravitational and gauge theoretic degrees of freedom, mediated by the $B$-field, leads to the Green-Schwarz anomaly cancellation mechanism~\cite{Green:1984sg}. Naively, one might expect a similar phenomenon to arise in the heterotic ambitwistor string, especially since coupling to the background gauge field (through the left-moving worldsheet current algebra) is the same as in string theory (cf., \cite{Moore:1984ws,Hull:1985jv}). This means that we should encounter the usual gauge anomaly associated with chiral fermions on the worldsheet.


In heterotic string theory, resolving this gauge anomaly relies crucially on the form of the $B$-field coupling to the worldsheet CFT (cf., \cite{Witten:1999eg}). This leads to a modified gauge transformation of the $B$-field and the shift of its field strength by a Chern-Simons term for the background gauge field. In a chiral model such as the ambitwistor string, the coupling of other background fields (including the $B$-field) is different and the resolution of the anomaly is no longer clear.

\medskip

To see this, it suffices to consider an abelian background gauge field $A_{\mu}$, now coupled to the worldsheet through an (abelian) current algebra of level $k\in\Z_{+}$. Under a gauge transformation $A_{\mu}\rightarrow A_{\mu}-\partial_{\mu}\varepsilon$, the worldsheet field $\Pi_{\mu}$ transforms as $\Pi_{\mu}\rightarrow\Pi_{\mu}-j\partial_{\mu}\varepsilon$. The gauge-transformed $\Pi_{\mu}$ now has a singular OPE with itself:
\be\label{ganom1}
(\Pi_{\mu}-j\,\partial_{\mu}\varepsilon)(z)\,(\Pi_{\nu}-j\,\partial_{\nu}\varepsilon)(w)\sim k\,\frac{\partial_{\mu}\varepsilon\,\partial_{\nu}\varepsilon}{(z-w)^2}-k\,\frac{\partial_{\mu}\varepsilon\,\partial(\partial_{\nu}\varepsilon)}{z-w}\,,
\ee
proportional to the level $k$. Such anomalous OPEs can be removed in chiral CFTs of $\beta\gamma$-type by compensating for the gauge transformation with a shift of the $\Pi_{\mu}$ field (e.g., \cite{Nekrasov:2005wg}). To remove the anomalous OPE \eqref{ganom1}, define the gauge transformation of $\Pi_{\mu}$ by:
\be\label{ganom2}
\Pi_{\mu}\rightarrow \Pi_{\mu}-j\partial_{\mu}\varepsilon +\frac{k}{2}\partial_\mu\varepsilon\,\partial\varepsilon\,.
\ee
It is straightforward to check that this level-dependent shift removes the singularities from \eqref{ganom1}, ensuring that $\Pi_{\mu}$ has a non-singular OPE with itself in any choice of gauge.

Unfortunately, there is still a gauge anomaly at the level of the worldsheet currents $\sG$ and $\sH$. It is straightforward to see that these currents are no longer gauge invariant; for instance
\be\label{ganom3}
\sG \rightarrow \sG +\frac{k}{2} \psi^{\mu} \partial_{\mu}\varepsilon\,\partial\varepsilon\,,
\ee
under a gauge transformation with the proviso \eqref{ganom2}. Furthermore, the OPE of $\sG$ with itself now has a triple pole contribution 
\be\label{ganom4}
\sG(z)\,\sG(w)\sim \frac{k\,A^{\mu} A_{\mu}}{(z-w)^3}+\cdots \,,
\ee
which must vanish. In other words, the $\sG$, $\sH$ current algebra imposes the gauge-dependent, algebraic equation of motion $A^2=0$ on the background gauge field.

A potential remedy for this situation would be to modify the current $\sG$. Such modification are constrained by conformal weight and fermionic statistics to take the form
\be\label{ganom5}
\sG\rightarrow \sG + \psi^{\mu}\psi^{\nu}\psi^{\rho}\,H_{\mu\nu\rho} + \psi^{\mu}\,C_{\mu\nu}\,\partial X^{\nu}\,,
\ee
for some $H_{\mu\nu\rho}$ and $C_{\mu\nu}$ that depend only on $X$. Corrections proportional to $H_{\mu\nu\rho}$ are reminiscent of the standard Green-Schwarz mechanism, but it is easy to see that they cannot remove the gauge-dependent triple pole \eqref{ganom4}. While we have been unable to use corrections proportional to $C_{\mu\nu}$ to remove the gauge anomaly completely, there are some suggestive hints which emerge.

Consider the modification of $\sG$ given by
\be\label{gmod1}
\sG\rightarrow \sG +\frac{k}{2}\,\psi^{\mu}\, A_{\mu}A_{\nu}\,\partial X^{\nu}\,.
\ee
It is worth noting that analogous terms appear in descriptions of anomaly cancellation for the Green-Schwarz formalism of the heterotic string~\cite{Hull:1986xn}, although the precise connection (if any) is unclear. This modification removes the triple pole \eqref{ganom4} entirely, and leads to $O(k)$ modifications of $\sH$:
\begin{multline}\label{hmod1}
 \sH\rightarrow \sH + k \left(\Pi^{\mu}\,A_{\mu}A_{\nu}\,\partial X^{\nu}-\psi^{\mu}\psi^{\nu}\,A_{\mu}\partial A_{\nu}-A^{2}\,A_{\mu}\partial X^{\mu}\,j  -\frac{1}{2} \partial A_{\mu}\,\partial A^{\mu} \right. \\
 \left. +\frac{1}{2} \partial\left(\partial_{\mu}(A^{\mu} A_{\nu})\partial X^{\nu}\right) - \psi^{\mu}\psi^{\nu}\,\partial_{\mu}(A_{\nu}A_{\sigma})\partial X^{\sigma}\right) +\frac{k^2}{4}\,A^2\,(A_{\mu}\partial X^{\mu})^2\,.
\end{multline}
The modified $\sG$ and $\sH$ are worldsheet conformal primaries of the appropriate dimension, although both are still gauge-dependent. 

While these modifications do not remove all gauge-dependence, the OPE of $\sG$ with $\sH$ takes a remarkably simple form. Indeed, on the support of the Maxwell equations for the background gauge field, one finds the structure:
\begin{multline}\label{ganom6}
 \sG(z)\,\sH(w)\sim k\left[\frac{\psi^\mu \psi^{\nu}\psi^{\rho}}{(z-w)^2} (\cdots) + \frac{\psi^{\mu}\psi^{\nu}\psi^{\rho} \partial X^{\sigma}}{z-w} (\cdots) + \frac{\psi^{\mu}\psi^{\nu}\partial\psi^{\rho}}{z-w} (\cdots) \right. \\ 
 \left. +\frac{\psi^{\mu}\,\partial X^{\nu}}{z-w} (\cdots) + \frac{\psi^{\mu}\partial X^{\nu} \partial X^{\rho}}{z-w} (\cdots) + \frac{\partial \psi^{\mu} \partial X^{\nu}}{z-w} (\cdots)\right]\,,
\end{multline}
where the $(\cdots)$ stand for gauge-dependent tensors constructed from the background gauge field. Although far from satisfactory, these modifications do kill all $O(k^2)$ contributions to the OPE, as well as terms proportional to $\Pi_{\mu}$ and $j$ -- all of which occur for generic modifications \eqref{ganom5}. Furthermore, many of the terms appearing in \eqref{ganom6} actually have a surprisingly simple form; for instance, the double pole is
\be\label{ganom7}
\sG(z)\,\sH(w)\sim -\frac{3k}{2}\,\frac{\psi^\mu \psi^{\nu}\psi^{\rho}}{(z-w)^2}\,A_{\mu} F_{\nu\rho} + \frac{1}{z-w} (\cdots)\,,
\ee
namely, a Chern-Simons term. 

We expect that a full resolution of the gauge anomaly for the heterotic ambitwistor string requires a full knowledge of its coupling to other background fields. These fields obey higher-derivative equations of motion, and there are additional fields (such as a massless 3-form) which do not appear in the standard heterotic string~\cite{Mason:2013sva,Azevedo:2017lkz,Azevedo:2017yjy}. A recent description of the effective free theory on space-time for these fields should be a useful tool in this regard~\cite{Berkovits:2018jvm}. We hope that future work will lead to a full resolution of these issues.

\acknowledgments

We would like to thank Carlo Meneghelli and David Skinner for useful discussions. TA is supported by an Imperial College Junior Research Fellowship; EC is supported by EPSRC grant EP/ M018911/1; SN is supported by EPSRC grant EP/M50659X/1 and a Studienstiftung des deutschen Volkes scholarship.

\bibliography{biblio}
\bibliographystyle{JHEP}

\end{document}